\documentclass[runningheads,orivec]{llncs}
\pdfoutput=1
\usepackage[T1]{fontenc}
\usepackage{graphicx}
\usepackage{hyperref}

\usepackage{color}

\usepackage{amsmath}
\usepackage{mleftright}
\usepackage{xspace}
\usepackage{mathtools}

\usepackage{wrapfig}

\usepackage{paralist}
\usepackage{tabularx}
\usepackage{multirow}
\usepackage{enumitem}

\usepackage{cleveref}
\Crefname{equation}{Formula}{Formulas}

\makeatletter
\AddToHook{cmd/appendix/before}{\def\cref@section@alias{appendix}}
\makeatother

\usepackage{todonotes}

\usepackage{TeX/lsyntax}
\usepackage{TeX/manalysis}
\usepackage[differentialdL,precisenames]{TeX/dL}

\usepackage{notation}
\usepackage{customMacros}

\usepackage{cite}

\usepackage{stmaryrd}

\begin{document}
\title{Verification of Autonomous Neural Car Control\\ with \KeYmaeraX}
\author{Enguerrand Prebet\orcidID{0009-0008-0160-5219} \and
Samuel Teuber\orcidID{0000-0001-7945-9110} \and
Andr\'e Platzer\orcidID{0000-0001-7238-5710}}%
\authorrunning{E. Prebet et al.}
\institute{Karlsruhe Institute of Technology, Karlsruhe, Germany
\email{\{enguerrand.prebet,teuber,platzer\}@kit.edu}}
\maketitle              %
\begin{abstract}

\looseness=-1
This article presents a formal model and formal safety proofs for the ABZ'25 case study in differential dynamic logic (\dL).
The case study considers an autonomous car driving on a highway with a neural network controller avoiding collisions with neighbouring cars.
Using \KeYmaeraX's \dL{} implementation we prove collision-freedom on an infinite time horizon which ensures that safety is preserved independently of trip length.
The safety guarantees hold for time-varying reaction time and brake force.
Our \dL model considers the single lane scenario with cars ahead or behind.
We demonstrate \dL and its tools are a rigorous foundation for runtime monitoring, shielding, and neural network verification.
Doing so sheds light on inconsistencies between the provided specification and simulation environment \texttt{highway-env} of the ABZ'25 study. %
We attempt to fix these inconsistencies and uncover numerous counterexamples %
indicative of
issues in the provided reinforcement learning environment.%
\end{abstract}

\section{Introduction}
\looseness=-1
This paper contributes a comprehensive study of formal safety proofs for the ABZ'25 highway case study of straight-line driving on highways with a neural network (NN) control system for the ego car based on the rigorous foundations of differential dynamic logic~\cite{DBLP:journals/jar/Platzer08,Platzer10,DBLP:journals/jar/Platzer17,Platzer18} (\dL).
Given the interest in highway driving, the contributions to the ABZ'25 case study challenge stand a more general appeal.
While the specific outcomes focus on the ABZ'25 case study, the generality of the underlying tools %
could help make other applications safe.

\paragraph{Contributions.}
\looseness=-1
To tackle ABZ's case study we provide: %
\begin{inparaenum}[\it i)]
    \item A formal, provably safe \dL~\cite{DBLP:journals/jar/Platzer08,Platzer10,DBLP:journals/jar/Platzer17,Platzer18} model of the hybrid systems dynamics of straight-line driving described by ABZ'25~\cite{ABZSpec}.
    We identify the control constraints required for safe driving.
    \item A derivation of real arithmetic constraints that serve either as sandbox/shield for the black-box NN or %
    for the \emph{gapless} rigorous white-box verification of concrete NNs.
    \item A verification-based, exhaustive characterization of all unsafe behaviours in two NNs trained using the \texttt{highway-env} environment provided by ABZ'25.
    \item An empirical validation of the derived sandbox and shield. %
\end{inparaenum}

\looseness=-1
Importantly, our safe %
controller and
the derived monitoring/verification conditions are fully symbolic and proved safe for arbitrary parameter choices
making the model, controller, sandbox and NN verification technique useful for future endeavours.
Additionally, reaction time and braking power may vary (within bounds) during execution.
The %
results underscore that safety guarantees in \dL are
practically applicable to (neural) real-world systems
-- either through monitoring/shielding or via verification of the NN
w.r.t.\ \dL derived constraints.

\looseness=-1
While we demonstrate that \dL and implementation monitoring/verification can be \emph{gaplessly} integrated,
we observe the existence of a significant \emph{model-to-simulation} (\emph{model2sim}) gap between the specification and the simulator provided by ABZ~\cite{ABZSpec}.
The well-known \emph{sim2real gap} leads to decreased performance when simulation-trained agents are deployed in the real world.
Similarly, the \emph{model2sim} gap induces unsafe behaviour of an agent if the simulation insufficiently matches the model's assumptions about the real world.
We identify this gap as an important roadblock on the highway to safe NN controllers. %

\paragraph{Related Work.} %
\looseness=-1
Prior work analysed safe car control in \dL~\cite{DBLP:conf/icfem/RenshawLP11,DBLP:journals/lmcs/Platzer12b,DBLP:conf/fm/LoosPN11} (in one instance using refinement~\cite{DBLP:conf/lics/LoosP16}).
Unlike prior case studies applying \dL guarantees to NN control~\cite{Fulton2018,TeuberVerSAILLE2024},
this work has a more complex environment (e.g.\ variable speed for surrounding cars) which increases the complexity of safety criteria.
Car control (with different dynamics~\cite{DBLP:journals/tecs/TranCLMJK19}) has also been studied by numerous closed-loop NN verification tools (see e.g.\ the ARCH competition~\cite{lopez2022arch,DBLP:conf/arch/LopezAFJL023,ARCH-COMP24:ARCH_COMP24_Category_Report}).
Unlike the closed-loop approaches, our work can provide guarantees on an infinite-time horizon, i.e.\ independent of the car's trip length.
Event-B~\cite{DBLP:books/daglib/0024570} has also been used to model automotive applications~\cite{DBLP:conf/ictac/BanachB13} without application to NNs.
Unlike a \texttt{highway-env} ProB model~\cite{DBLP:conf/nfm/VuDL24} we explicitly model the environment's continuous dynamics and support NN verification.
Unlike another shielding approach~\cite{DBLP:conf/collas/Shperberg0AS22} we characterize safe behaviour a priori instead of learning from catastrophic behaviour.

\section{Background} %
\looseness=-1
This section provides an overview of differential dynamic logic (\dL).
Before presenting results on highway car control,
we first illustrate the concepts of this section using a cartoonishly simplified application:
We consider a car that starts at a one-dimensional, positive position $p$ %
and pretend
the car's controller can influence the car's position by directly choosing the car's velocity $v$ with immediate effect.
The safety requirement of the controller is to keep the car at a positive position, i.e.\ $p>0$.
We first present \dL in general, then the ModelPlex technology for the derivation of runtime monitors and three applications of these formulas.

\subsection{Differential Dynamic Logic for Hybrid Systems}
\label{subsec:background:dl}
\dL is a program logic for reasoning about cyber-physical systems given as \emph{hybrid programs}.
On a high level, \dL is a first-order multi-modal logic where modalities are parameterized with programs and the first-order formulas are interpreted w.r.t.\ real arithmetic.
Formulas of \dL have the following structure:

\begin{definition}[Formulas]
	\emph{Formulas} are defined by the grammar below where $\astrm,\bstrm$ are terms, $\asfml,\bsfml$ are formulas and $\asprg, \bsprg$ are hybrid programs (\Cref{def:HP}):
\[\begin{array}{c c c}
    \asfml, \bsfml & \Coloneqq & \astrm \leq \bstrm \OR \lnot \asfml \OR \asfml \land \bsfml \OR \lforall{x}{\asfml} \OR \dbox{\asprg}{\asfml} \OR \asprg \refines \bsprg
\end{array}\]
\end{definition}
\looseness-1
While the first four elements of the grammar correspond to logical structures known from first-order real arithmetic formulas, the latter two are specific to differential dynamic logic~\cite{DBLP:journals/jar/Platzer08,Platzer10,DBLP:journals/jar/Platzer17,Platzer18} and differential refinement logic~\cite{DBLP:conf/lics/LoosP16,DBLP:conf/ijcar/PrebetP24}.

Unlike first-order formulas which are usually evaluated in a fixed structure, \dL evaluates formulas w.r.t.\ states that assign values to variables.
The programs (which will be discussed in greater detail below) then induce a state transition relation which is integrated into the logic via the grammar's fifth formula:
$\dbox{\asprg}{\asfml}$ is true in a state $\omega$ iff \emph{after every program run} of $\asprg$ the formula $\asfml$ is satisfied, i.e.\ if for all state transitions of $\asprg$ from the current state $\omega$ the formula $\asfml$ holds in the resulting state.
If $\asfml$ is a property that indicates safety of the system, then $\dbox{\asprg}{\asfml}$ expresses that the system always remains safe (see \Cref{sec:model:proof}).
Finally, $\asprg \refines \bsprg$
expresses that the program $\asprg$ \emph{refines} the program $\bsprg$ in the current state, i.e.\ $\asprg \refines \bsprg$ holds in a state $\omega$ iff all states reachable from $\omega$ via the transitions of $\asprg$ are also reachable via $\bsprg$'s state transition relation.
Refinements are used to transfer safety properties between hybrid programs (see \Cref{subsec:safe_nn:refinement}).
A formula is called \emph{valid} if it is satisfied in all states.
We now turn to \dL's \emph{hybrid programs} which allow discrete and continuous actions and are formally defined as follows:

\begin{definition}[Hybrid Programs]\label{def:HP}
	\emph{Hybrid programs} $\asprg,\bsprg$ are defined by the grammar below where $x$ is a variable, $\astrm$ is a term and $\bsfml$ is a formula:
\[\begin{array}{c c c}
	\asprg, \bsprg & \Coloneqq & \ptest{\bsfml} \OR \pupdate{\pumod{x}{\astrm}} \OR \prandom{x} \OR \pode{\D{x}=\astrm}{\bsfml} \OR \pchoice{\asprg}{\bsprg} \OR \asprg;\bsprg \OR \prepeat{\asprg}
\end{array}\]
\end{definition}
The first program primitive $\ptest{\bsfml}$ (check) only proceeds if formula $\bsfml$ is satisfied in the current state.
The second and third primitive are assignments, either w.r.t.\ a term ($\pupdate{\pumod{x}{\astrm}}$) or nondeterministically to an arbitrary value ($\prandom{x}$).
The fourth primitive ($\pode{\D{x}=\astrm}{\bsfml}$) describes the continuous, nondeterministic evolution of variable $x$ along the differential equation $\D{x}=\astrm$  within the domain constraint $\bsfml$.
The next two primitives allow the composition of programs by either nondetermnistically choosing one of two ($\pchoice{\asprg}{\bsprg}$) or by executing them sequentially ($\asprg;\bsprg$).
The final primitive $\prepeat{\asprg}$ nondeterministically runs the program $\asprg$ for 0 or more iterations.
The support of hybrid programs for continuous evolution and discrete as well as continuous nondeterminism is e.g.\ crucial for the analysis of cyber-physical systems without a fixed clock cycle.
Many classical program constructs can be translated into the primitives of hybrid programs.
For example, if-then-else can be rewritten as follows:
$
\texttt{if}\left(\bsfml\right)~\asprg~\texttt{else}~\bsprg
\enspace\defeq\enspace
\pchoice{\left(\ptest{\bsfml};\asprg\right)}{\left(\ptest{\neg\bsfml}\right);\bsprg}
$.
Similarly, we can represent while loops:
$
\texttt{while}\left(\bsfml\right)~\asprg
\enspace\defeq\enspace
\prepeat{\left(\ptest{\bsfml};\asprg\right)};\ptest{\neg\bsfml}
$.

\paragraph{Example.}
We now explain how our simple cyber-physical system (the velocity-controlled car) can be modelled in \dL.
All variables, \(p,v,\dots\), that evolve along the execution are in lower-case, while constants like $T$ are in upper-case.
As outlined above, the car's position is described by a real-valued position $p$.
The car's dynamics are then described by the differential equation $\D{p}=v$ where $v$ is the velocity determined by the controller.
To derive safety guarantees we assume that our controller is invoked at least every $T$ seconds.
Hence, we model the physical part of our example as 
$
\alpha_{\text{plant}} \enspace\defeq\enspace \pupdate{\pumod{t}{0}};\,\pode{\D{p}=v,\D{t}=1}{t\leq T}
$.
Here, $p$ evolves as outlined above and we additionally introduced a clock variable $t$ which guarantees that the evolution runs for at most $T$ seconds via the domain constraint $t \leq T$.
We already formulated the car's safety condition as $p>0$ at the beginning of this section.
The final ingredient for our \dL model is a \emph{control envelope} that provides a nondeterministic description of allowed behaviour which keeps the system safe.
Using \dL to verify control \emph{envelopes} rather than one concrete controller is quite a common approach as it allows the verification of a whole family of possible controller implementations at once \cite{DBLP:conf/tacas/KabraLMP24}.
It is generally preferable to design very general control envelopes that encompass the largest possible range of behaviours that can be certified as safe.
In our example, we can formulate the control envelope as the nondeterministic program
$
\alpha_{\text{ctrl}} \enspace\defeq\enspace
{\prandom{v};}\,{\ptest{\left(p+Tv > 0\right)}}
$.
This control envelope ensures that we only choose velocities $v$ that avoid negative positions.
Indeed, we can use \dL's proof calculus (and its implementation in \KeYmaeraX) to prove the validity of the following \dL formula
$
T>0 \land p>0 \rightarrow \dbox{\prepeat{\left(\alpha_{\text{ctrl}};\alpha_{\text{plant}}\right)}}{p>0}
$.
This formula expresses that (assuming an initial state with $T>0$ and $p>0$) we can run this system for arbitrarily long (note the nondeterministic loop) and the safety condition $p>0$ will always be satisfied afterwards.
This can be proven inductively through the invariant ${T>0}\,\land\,{p>0}$.
The ability to perform inductive, infinite time horizon reasoning for \dL models is one of the major advantages of \dL over many reachability-based analyses.

The formula above also exhibits a very common pattern in \dL models where we provide a safety guarantee over the execution of a nondeterministic loop which consists of the sequential execution of a control envelope ($\alpha_{\text{ctrl}}$) and an environment model ($\alpha_{\text{plant}}$).
However, while we have verified an infinite class of potential controllers, we have not yet verified any concrete given controller implementation.
In the remainder of this section, we will present \dL-based technologies that allow us to bridge the gap between a verified control envelope and a concrete controller implementation.

\subsection{ModelPlex for Verified Runtime Monitoring}
\label{subsec:background:modelplex}
In the previous section, we saw how \dL can be used to model cyber-physical systems and to verify control envelopes.
However, the verified control envelopes differ from the control systems we would like to use in practice:
Concrete, real-world controllers will often be implemented in compilable programming languages or, as in the case of the highway case study, the controller's behaviour might even be determined by an NN.
This raises the question how this challenge can be overcome.
On the one hand, it is possible to embed the behaviour of more complicated programming languages into \dL~\cite{Kamburjan2022,DBLP:conf/iccps/GarciaMP19}, however, such approaches are always tailored to specific programming languages and require that we perform interactive proofs on the concrete controller's behaviour.
On the other hand, we can use a verified control envelope to derive runtime monitoring conditions that can subsequently be checked on a concrete system -- possibly even in a black box fashion.
This technique to derive correct-by-construction runtime monitoring conditions from a given control envelope $\alpha_ {\text{ctrl}}$ is called \emph{ModelPlex}~\cite{Mitsch2016}.

\looseness=-1
Based on a given control envelope $\alpha_ {\text{ctrl}}$ over variables $\vars{\alpha_ {\text{ctrl}}}$, ModelPlex uses \dL's calculus rules to derive a first-order real arithmetic formula $\bsfml$ over variables
$
\vars{\alpha_ {\text{ctrl}}}
\dot{\cup}
\left\{
x^+ \middle| x\in\vars{\alpha_ {\text{ctrl}}}
\right\}
$ where $x^+$ indicates the value of $x$ in the next state.
For instance assuming $\vars{\alpha_ {\text{ctrl}}} = \left\{x\right\}$, if the formula $\bsfml$ is satisfied by $x=v_1$ and $x^+=v_2$ for some $v_1,v_2$, then
there exists a state transition for $\alpha_ {\text{ctrl}}$ where the value of $x$ changes from $v_1$ to $v_2$.
Since we have a safety proof for $\alpha_{\text{ctrl}}$ this implies
the safety guarantees for our control envelope carry over to a system where $x$'s value changes from $v_1$ to $v_2$.
Hence, the formula $\bsfml$ can be used to monitor the safety of a (black-box) controller implementation by checking whether a concrete assignment of the implementation's pre- and post-values satisfies $\bsfml$.

For the velocity-controller car, the variables $\vars{\alpha_{\text{ctrl}}}$ are the position and the velocity: $\{p,v\}$, and the formula given by ModelPlex is \(\bsfml \defeq {p^+ = p} \land {p + T v^+ > 0}\,\land\, {t^+=0}\).
Thus, any concrete implementation of such a controller will be safe if this formula is satisfied during execution, i.e\ if the controller does not change the position ($p = p^+$) and sets some velocity $v^+$ that respects \(p + T v^+ > 0\).
Additionally, it requires that the clock variable $t$ be reset to $0$.

\subsection{Applications of ModelPlex}
\label{subsec:background:applications}
\looseness=-1
The formula computed by ModelPlex~\cite{Mitsch2016} tells us which control actions come with a \dL0 safety guarantee.
As explained below, this formula can be used in at least three manners to derive safety guarantees for controller implementations.

\paragraph{Monitoring (VeriPhy).}
First and foremost, we can use the derived formula to check the actions computed by the controller implementation at runtime via a runtime monitor.
To this end, we assign the formula's variables with the implementation's input and output values and check whether the action is provably safe according to the ModelPlex runtime monitor.
In case the implementation chooses an action violating the runtime monitor,
we overwrite the action using a fallback controller.
This approach comes with a formally verified code generation pipeline called VeriPhy~\cite{Bohrer2018} which serves as a sandbox for a given controller and comprises provably correct machine arithmetic.

\paragraph{Shielding (Justified Speculative Control).}
\looseness=-1
One drawback of VeriPhy in the context of NN Control is its conservatism:
While traditionally programmed controllers usually return exactly one action that must be overwritten if unsafe, NNs often return a probability distribution over actions.
However, it is not necessarily reasonable to entirely overwrite the NN's action if its most likely action is unsafe.
Instead Justified Speculative Control~\cite{Fulton2018,DBLP:conf/tacas/FultonP19} (JSC) \emph{shields} the NN 
using runtime enforcement technique~\cite{DBLP:conf/birthday/KonighoferBEP22,Mitsch2016} that constrain the action space to known-safe options.
Thus, JSC can still treat the concrete controller as a black box but allow for more flexibility in the chosen actions.
To this end, JSC checks for possible actions whether they satisfy the ModelPlex condition.
JSC then chooses the allowed action with the highest probability according to the reinforcement learning agent.
Additionally, JSC only performs a safety check in situations where the environment behaves as modelled in \dL (this is achieved via ModelPlex's environment monitoring technology which goes beyond the scope of this exposition).
Importantly, this technique can be applied both during training and at runtime. %

\paragraph{Verification (VerSAILLE \& NCubeV).}
\looseness=-1
The previous approaches only provide \emph{a posteriori} guarantees by restricting or overwriting the controller's actions at runtime.
Alternatively, we can also use the monitoring condition derived by ModelPlex for \emph{a priori} verification of the NN.
This is achieved via the VerSAILLE approach~\cite{TeuberVerSAILLE2024}:
In essence, we verify whether there exists a state inside the \dL model's invariant state space 
where
the NN's action violates the ModelPlex controller monitor.
For this section's running example, we would verify that an NN (with input $p$ and output $v^+$) satisfies the following specification~\cite[Thm. 2]{TeuberVerSAILLE2024}:
\[
\underbrace{p>0}_{\text{Invariant}} \rightarrow \underbrace{p+T v^+ > 0}_{\text{Controller Monitor}}.
\]
\looseness=-1
This is achieved by a compute-intensive numerical analysis of the NN that mathematically proves the absence of such counterexamples.
As our running example has a simple, linear controller monitor and invariant, 
most modern NN verifiers (as reported in recent surveys and competitions~\cite{DBLP:journals/jmlr/0005BHR24,DBLP:journals/sttt/BrixMBJL23,DBLP:journals/corr/abs-2412-19985}) can be used. 
However, for realistic \dL models, the ModelPlex conditions usually have a significantly more complicated propositional structure with nonlinear real arithmetic.
Neither of these features is supported by ``classical'' NN verifiers nor by their common specification language%
~\cite{DBLP:conf/cav/DemarchiGPT23}.
To this end, we recently proposed the NCubeV tool~\cite{TeuberVerSAILLE2024} 
supporting
both arbitrary propositional structure and polynomial arithmetic.

\looseness=-1
The usage of NN verification has multiple advantages.
First, it allows the deployment of autonomous, unmonitored NN Control Systems.
Second, it allows the usage of NNs in applications without an obvious fallback strategy or for cases with continuous action spaces.
Finally, it can also serve for diagnostics:
Either to estimate how often a given NN performs (un)safe actions or to discover unsafe behaviour that is empirically invisible, e.g.\ due to simulator limitations.

\section{A Verified \dL Model for the ABZ Highway Case-Study}
\label{sec:model}
This section presents the verified \dL model developed for this case study.
We start by introducing the cyber-physical system of interest (\Cref{sec:model:intro}).
After giving the general structure and how it interleaves the discrete and continuous actions that can occur between each control cycle (\Cref{sec:model:struct}), we focus on the plant (\Cref{sec:model:plant}) and the controller (\Cref{sec:model:ctrl}).
Finally, we express safety conditions in \dL for the model and verify them using the theorem prover \KeYmaeraX \cite{DBLP:journals/jar/Platzer17,DBLP:conf/cade/FultonMQVP15,stefan_mitsch_2024_13380145} (\Cref{sec:model:proof}).
Our proofs are reproducible via an artifact~\cite{zenodoArtifact}.

\subsection{A Safe Autonomous Driving System}
\label{sec:model:intro}
The model is about a safe autonomous driving system, referred to as the ego car, that should prevent collision with another car on a single straight lane.
All constants, $\Amax,V,T,\dots$, must be positive except for braking deceleration, $\Bmin,\Bmax$, which are negative.
Both cars have length $L$ but are modelled as single points: with position $\xl$, speed $\vl$, and acceleration $\al$ for the ego car, and with position $\xf$, speed $\vf$, and acceleration $\af$ for the other.
Thus, absence of collision is ensured by maintaining a distance of at least $L$ between the two cars.
No car moves backwards and their speed is at most $V$. 
The cars have a maximum acceleration of $\Amax$ and a maximum braking deceleration of $\Bmax$.
Additionally, the ego car may not always draw the maximum power of the brake or the engine.
It will however always be able to brake with deceleration at least \(\Bmin \geq \Bmax\) and accelerate with acceleration at least \(\Amin \leq \Amax\).
These constraints are imposed on the cars themselves, so even if they are trying to brake or accelerate, they cannot go backward or exceed speed limit $V$.
The ego car observes the environment at least every $T$ seconds, whereas the other car may react more often without restriction.
No regularity or periodicity is assumed in the reaction time of the ego car as long as it always remains below $T$ seconds.

Overall, the constants are constrained by the formula:
\(
\constctx \defeq {T > 0} \land {L > 0} \,\land\, {V > 0}\, \land\, {\Bmax \leq \Bmin < 0 < \Amin \leq \Amax}
\).
It can be extended by bounds on speed and acceleration: \(\ctxt\defeq \constctx \,\land \,{\Bmax \leq \af,\al \leq \Amax}\, \land\, {0 \leq \vf,\vl \leq V}\).
\subsection{Overall Structure of the \dL Model}
\label{sec:model:struct}
The general structure of the model is as follows:
\[\model(\texttt{c}) \Coloneqq \prepeat{\big(\underbrace{\ctrlU;(\pchoice{\texttt{c}}{\ptest{t < \clock + T}})}_{\text{control}};\underbrace{\accCor;\plantT}_{\text{plant}}\big)}\]
The model is parametric in the controller of the ego car \texttt{c} to handle both the generic controller $\ctrlT$ (see \Cref{sec:model:ctrl}) and the NN controller $\ctrlNN$ (see \Cref{subsec:safe_nn:refinement}).
In this section, we write $\model$ for $\model(\ctrlT)$.

\ctrlU{} models the controller of the other car. It does not assume any minimal time between each execution of \ctrlU{}.
Then \texttt{c} models the controller of the ego car and sets $\clock$ to $t$.
If it has been less than $T$ seconds since $\clock$ was last set, the nondeterministic choice allows \texttt{c} to be skipped. Thus, the controller is only assumed to run at least once every $T$ seconds.
Having the possibility of skipping the controller allows discrete events, e.g.\ the other controller, to still occur without the ego car reacting.
\accCor{} (defined in \Cref{sec:model:plant}) models the acceleration correction when reaching the speed boundaries.
It ensures that a braking car, with negative acceleration, does not go backwards by changing its acceleration to zero.
This is a discrete change but happens independently of any controller. In particular, the ego car does not notice the change before its next control cycle.
Finally, \plantT{} models the continuous dynamics of the system, i.e.\ the actual motion of the car evolving with time.
These execute in a loop so that the system alternates between the control and the plant arbitrarily many times.
We elaborate the details of each component, starting with the plant.

\subsection{Modelling the Physical Plant}
\label{sec:model:plant}
\begin{tabular}{r|l}
\begin{minipage}{0.2\textwidth}
\begin{flushright}
\accCor\\[0.5em]
~\\
\dyn\\
~
\end{flushright}
\end{minipage}
&
\begin{minipage}{0.4\textwidth}
\begin{tabbing}
\texttt{if} \((\vl = 0 \land \al < 0)\lor(\vl = V \land \al > 0)\,\pupdate{\pumod{\al}{0}}\)\\
\texttt{if} \((\vf = 0 \land \af < 0)\lor(\vf = V \land \af > 0)\,\pupdate{\pumod{\af}{0}}\)\\[0.5em]
\(\D{\xf}=\vf,\)\=\(\D{\vf}=\af,\D{\xl}=\vl,\D{\vl}=\al,\D{t}=1\)\\
\>\(\& ~t \leq \clock + T \land 0 \leq \vf \leq V \land 0 \leq \vl \leq V\)
\end{tabbing}
\end{minipage}
\end{tabular}

The plant is composed of a discrete part, \accCor, and a continuous part, \plantT.
First, if any car has come to a stop or reached their speed limit, then their acceleration is set to 0 for saturation.
Then the continuous dynamics follows the ODEs specifying for both cars, that speed is the derivative of the position, \(\D{\xg}=\vg\), and that acceleration is the derivative of speed, \(\D{\vg}=\ag\).
Time is explicit with constant derivative.
The domain constraints ensure that the dynamics always stop before a discrete event must be executed, whether it is a controller event -- if \(t = \clock + T\) -- or a plant event -- if a car stops, or reaches their speed limit.

\subsection{Modelling the Car Controllers}
\label{sec:model:ctrl}
\begin{wraptable}[8]{r}{0.58\linewidth}
\vspace*{-1cm}
\begin{tabular}{r|l}
\begin{minipage}{0.12\linewidth}
\begin{flushright}
\ctrlU\\[.5em]
\ctrlT\\[.7em]
~\\~\\~\\~\\~
\end{flushright}%
\end{minipage}
&
\begin{minipage}{0.4\linewidth}
\begin{tabbing}
    \(\prandom{\al};\ptest{(\Bmax \leq \al \leq \Amax)};\)\\[.5em]
    \(\prandom{\af};\ptest{(\Bmax \leq \af \leq \Amax)};\pumod{\clock}{t};\)\\
    \texttt{if}\= \((\lnot(\safeBackT\lor\safeFrontT))\)\\
    \> \texttt{if}\= \((\xf \leq \xl)\)\\
    \> \> \(\prandom{\af};\ptest{(\Bmax \leq \af \leq \Bmin)};\)\\
    \> \texttt{else}\\
    \> \> \(\prandom{\af};\ptest{(\Amin \leq \af \leq \Amax)};\)
\end{tabbing}
\end{minipage}
\end{tabular}
\end{wraptable}
\looseness-1
The control consists of the controllers for the two cars.
The controller $\ctrlU$ for the other car isn't concerned about safety so it just selects any acceleration within the limitation of the vehicle.
As the assignment is nondeterministic, all choices of acceleration are taken into account for the safety proof. 
Then the controller for the ego car also selects an arbitrary acceleration. It however performs an additional check.
If the chosen acceleration does not satisfy one of the safety conditions, \safeBackT{} or \safeFrontT{} discussed below, then a fallback procedure overrides the acceleration.
The fallback simply tries to increase the distance with the other car. If the ego is behind, it brakes with \(\af \leq \Bmin\), and accelerates, \(\af \geq \Amin\), if ahead.
Finally, $\clock$ is set to $t$ to record the last time the controller ran.

\paragraph{Safety condition when behind.}
We focus on $\safeBackT$ shown in \Cref{eq:safeback}.
It expresses when an acceleration guarantees safety when the ego car is behind the other car.
First, the two cars should be at distance at least $L$ from each other, as that would correspond to a collision otherwise.
Additionally, if both cars were to brake, there should still be a distance at least $L$ when they stop.
For a braking ego car with acceleration \(\af < 0\), it stops at position \(\dist_\indEgo(\af) \defeq \xf - \frac{\vf^2}{2\af}\) meters.
For the other car, we assume the worst case. This happens when the other car's acceleration is directed towards the ego car, that is when it is braking at maximum force, \(\al = \Bmax\), in which case it stops at position \(\dist_\indOther \defeq \xl - \frac{\vl^2}{2\Bmax}\).
With constant acceleration, if the current position of the cars and their stopping position are both at safe distance, then these properties are invariants of the dynamics and thus ensure collision-freedom.
Changing acceleration for the other car can only increases its distance to the ego car and so does not risk collision.
\begin{alignat}{2}
    \label{eq:safeback}
    \xf &+ L \leq \xl \land \big(\af \leq \Bmin \land \dist_\indEgo(\Bmin) + L < \dist_\indOther \nonumber\\
    &\lor {\Bmin} \leq \af \land \vf + \af T < 0 \land \dist_\indEgo(\af) + L < \dist_\indOther\\
    &\lor  {\Bmin} \leq \af \land \vf + \af T \geq 0 \land \dist_\indEgo(\Bmin) + \corr + L < \dist_\indOther\big) \nonumber
\end{alignat}

To handle the ego car's change of acceleration, this idea is refined further and split in three scenarios:
\begin{enumerate}
    \item Since the ego car is only assumed to be able to brake with \(\af = \Bmin\) for sure, even if it is currently braking more,
    we still must rely on the minimum braking deceleration for checking the distance, so we use \(\dist_\indEgo(\Bmin)\).
    \item If \(\af \geq \Bmin\) but the car will stop before $T$ seconds, then the acceleration $\af$ can be used directy. Once stopped, the car remains safe, so we use \(\dist_\indEgo(\af)\).
    \item Otherwise, we must check that the car can start braking at the next control cycle, after at most $T$ seconds, and stop before crashing. This reuses the first case, with a correction term to account for the distance travelled and the speed change before the next cycle: \(\corr \defeq (\frac{-\af}{\Bmin} + 1)(\frac{\af}{2}T^2 + T\vf)\).
\end{enumerate}

\paragraph{Safety condition when ahead.}
\looseness-1
If the ego car is ahead, the setting is similar when changing the frame of reference.
From the perspective of an observer moving at constant speed $V$, the two cars are moving at speed $\fram{v}_{\ind}\defeq \vg-V$ in the opposite direction.
Their positions are now \(\fram{x}_{\ind}\defeq\xg-V\times t\),
and the worst case occurs when the other car approaches the ego car with maximal acceleration (i.e.\ \(\al = \Amax\)).
Reusing the insight for the previous case, we consider their stopping position in that new frame of reference (\(\fram{v}_{\ind}=0\)), which amounts to reaching maximum speed (\(\vg = V\)).
This gives the following distances updated with the new variables: \(\fram{\dist}_\indEgo(\af) \defeq \fram{x}_{\indEgo} - \frac{\fram{v}_{\indEgo}^2}{2\af}\) for the ego car, and \(\fram{\dist}_\indOther \defeq \fram{x}_{\indOther} - \frac{\fram{v}_{\indOther}^2}{2\Amax}\) for the other.
The resulting formula $\safeFrontT$ is given in \Cref{app:fig}.

\subsection{Safety Proofs}
\label{sec:model:proof}
Now that the model is defined comes the actual verification.
Since the goal is to prevent collisions, the safety condition is simply that the two cars have at least a distance $L$ between them.
Being on a single lane, they cannot cross each other, so the order of the cars remains the same, so the two cases when the ego car is behind or ahead can be proved independently.
Due to their similarity, we again focus on the case where the ego car is behind.
The general assumptions include the constraints from the specifications from \Cref{sec:model:intro}, i.e.\ $\ctxt$, and assume the controller of the ego car has last been run $T$ seconds ago so that it must run initially, i.e.\ \(\clock = t-T\).
The only other requirement is that initial states where a crash is unavoidable are prohibited, in which case, no controller can guarantee safety.
This the initial condition correspond to the first case of \Cref{eq:safeback}: the fallback action should give enough distance before stopping.

\begin{theorem}
    \label{thm:model:dl_safety}
    \Cref{eq:safety:back,eq:safety:front} are valid and guarantee absence of collision.
    \begin{align}
\label{eq:safety:back}
\ctxt \land \xf + L \leq \xl \land \dist_\indEgo(\Bmin) + L < \dist_\indOther \land \clock = t-T &\limply \dbox{\model}\,\xf + L \leq \xl
\\
\label{eq:safety:front}
\ctxt \land \xl + L \leq \xf \land \fram{\dist}_\indOther + L < \fram{\dist}_\indEgo(\Amin) \land \clock = t-T &\limply \dbox{\model}\,\xl + L \leq \xf
\end{align}
\end{theorem}
The theorem is proved using \KeYmaeraX.
The proof relies on invariants that generalise of $\safeBackT$ and $\safeFrontT$ where $T$ is replaced by \(T + \clock-t\) to account for the time elapsed since the last run of $\ctrlT$, extended with the specification constraints $\ctxt$.
The evaluation of the two verifications is given in \Cref{app:fig}.

\section{Safeguarding Neural Control}
\label{sec:safe_nn}
The previous section derived a \dL model for the highway environment as specified in ABZ's case study document~\cite{ABZSpec} and proved its safety.
As a next step, we connect these (abstract) safety guarantees to the concrete control system implementation running inside the \texttt{highway-env} simulation~\cite{highway-env}.
To this end, we use the techniques described in \Cref{subsec:background:applications}.
In contrast to the \dL controller $\ctrlT$ that chooses a (continuous) acceleration value $\Bmax\leq\af\leq\Amax$, the trained reinforcement learning agent for the single-lane case of \texttt{highway-env} consists of an NN outputting one of three discrete actions (brake, idle, accelerate).
The NN outputs three values and determines its action via an argmax operation (e.g.\ brake is chosen whenever the NN's first output is maximal), prioritising lowest speed in case of ties.
Hence, we must first extend our \dL controller model to account for the NN's three outputs (\Cref{subsec:safe_nn:refinement}).
Subsequently, we can use ModelPlex and the refined controller to derive a formula that can be used for verification, shielding and monitoring (\Cref{subsec:safe_nn:modelplex}).
While our methodology is general, this section focuses on the case where the ego car drives behind another car and must ensure safety.

\subsection{Refining the \dL Controller}
\label{subsec:safe_nn:refinement}
To account for the concrete NN, we transform the controller's action space from choosing an acceleration $\af$ to choosing an action via three outputs $\yOne,\yTwo,\yThree$:
\begin{tabular}{r|l}%
\begin{minipage}{0.2\textwidth}%
\begin{flushright}%
\ctrlNN\\%
~\\~\\~\\~\\~\\~%
\end{flushright}%
\end{minipage}%
&%
\begin{minipage}{0.4\textwidth}%
\begin{tabbing}
    \(\prandom{\yOne};\prandom{\yTwo};\prandom{\yThree};\)\\
    \texttt{if}\= \((\yOne \geq \yTwo \land \yOne \geq \yThree)\enspace\{\prandom{\af};\ptest{(\Bmax \leq \af \leq \Bmin)}\}\);\\
    \texttt{if}\= \((\yTwo > \yOne \land \yTwo \geq \yThree)\enspace\{\pupdate{\pumod{\af}{0}}\};\)\\
    \texttt{if}\= \((\yThree > \yOne \land \yThree > \yTwo)\enspace\{\prandom{\af};\ptest{(\Amin \leq \af \leq \Amax)}\};\)\\
    $\{$\hphantom{$\cup$}        \>\enspace  \(?(\xf \leq \xl \land \Bmax \leq \af \leq \Bmin)\)\\
    \hphantom{$\}$}$\cup$    \> \enspace \(?(\xf \geq \xl \land \Amin \leq \af \leq \Amax)\)\\
    \hphantom{$\}$}$\cup$    \>\enspace  \(?(\safeBackT\lor\safeFrontT)\enspace\};\pupdate{\pumod{\clock}{t}}\)
\end{tabbing}%
\end{minipage}%
\end{tabular}\\
\looseness=-1
Based on the NN's outputs $\yOne,\yTwo,\yThree$ the program determines the corresponding acceleration value $\af$ and then ensures safety via the checks we already know from the \dL model for $\ctrlT$.
To recover the formal guarantee from \Cref{eq:safety:back}, we show that $\modelNN$ refines $\model(\ctrlT)$, i.e.\ $\modelNN$'s transitions are included in $\model(\ctrlT)$'s.
In fact, we prove a slightly relaxed refinement to ignore the variables $\yOne,\yTwo,\yThree$ that are modified by $\ctrlNN$ and not $\ctrlT$.

\begin{lemma}
The following refinement is valid:
\begin{equation}\label{eq:refinement}
    \constctx\enspace\rightarrow\enspace\left(\enspace\modelNN \refines \left(\prandom{\yOne};\prandom{\yTwo};\prandom{\yThree};\model(\ctrlT)\right)\enspace\right)
\end{equation}
\end{lemma}
Using the refinement, it is then trivial to extend the proof of \Cref{eq:safety:back} to $\modelNN$.
The proof of refinement is done using \KeYmaeraX's differential refinement logic implementation\footnote{\url{https://github.com/LS-Lab/KeYmaeraX-release/tree/dRL-ABZ'25}} and is based on a proof of refinement between $\ctrlNN$ and \(\prandom{\yOne};\prandom{\yTwo};\prandom{\yThree};\ctrlT\).

\subsection{ModelPlex for Safe Neural Network Control}
\label{subsec:safe_nn:modelplex}
We have now shown that any action taken by $\ctrlNN$ keeps the system safe on an infinite-time horizon.
Using ModelPlex we derive a controller monitor for $\ctrlNN$ that we can use w.r.t.\ a concrete NN.
To this end, we note that according to the specification~\cite{ABZSpec} the NN has (among other inputs) a vector of inputs $=\left(\xf,\vf,\xl,\vl\right)$ and the NN's only output is a vector $\overline{\texttt{out}}=\left(\yOne^+,\yTwo^+,\yThree^+\right)$.
Besides the variables in \nnInputVec{}, \nnOutputVec{} the controller monitor derived via ModelPlex also constrains the acceleration variables $\af$ and $\af^+$ (as \ctrlNN{} modifies $\af$) as well as the clock variables $t,\clock^+$ (required for book-keeping on control cycles).
We denote this ModelPlex condition for \ctrlNN{} as $\modelplexCondOrigInOut$.
As described in \Cref{subsec:background:applications}, VerSAILLE allows us to use the monitor $\modelplexCondOrigSym$ to verify the safety of an NN by additionally exploiting the \dL model's loop invariant which tells us what states are reachable (and thus for which states the NN must exhibit safe actions).
We denote this invariant as $\invariantCondOrigIn$.
In addition to the two cars' positions and velocities, the invariant also mentions the cars' accelerations and the clock variables $t,\clock$.
As explained in \Cref{subsec:background:applications} we can prove the infinite-time horizon safety of an NN by showing that all inputs inside the invariant $\invariantCondOrigSym$ lead to outputs satisfying the controller monitor $\modelplexCondOrigSym$.
Formally, this can be expressed as the following Theorem which follows from \cite[Thm. 2]{TeuberVerSAILLE2024}:

\begin{theorem}[NNCS Safety Criterion]
Let $g$ be an NN for highway car control as modeled in \Cref{sec:model}.
If \Cref{eq:safe_nn:versaille_condition} is satisfied for all \nnInputVec{} and $\nnOutputVec = g\left(\nnInputVec\right)$ then the safety guarantees derived in \Cref{thm:model:dl_safety} apply to $\modelg$.
\begin{equation}
    \label{eq:safe_nn:versaille_condition}
    \forall \af,\af^+,\al,t,\clock,\clock^+\enspace
    \underbrace{\invariantCondOrigIn}_{\text{system invariant}} \rightarrow \underbrace{\modelplexCondOrigInOut}_{\text{monitoring formula}}
\end{equation}
\end{theorem}
While this work omits the precise formulation, it is worth noting that the safety guarantees for $g$ are rigorously founded in \dL via a reconstruction of $g$ inside \dL through the notion of \emph{nondeterministic mirrors}~\cite[Def. 16]{TeuberVerSAILLE2024}.

\begin{wrapfigure}[9]{r}{0.6\textwidth}
    \vspace*{-.1cm}
    \begin{align*}
    \omit{\rlap{$\modelplexCtx~\defeq$}}\\
    \yOne^+ \geq \yTwo^+ \land \yOne^+ \geq \yThree^+ & \rightarrow \Bmax \leq \af^+ \leq \Bmin~\land \\
    \yTwo^+ > \yOne^+ \land \yTwo^+ \geq \yThree^+ & \rightarrow \af^+ = 0~\land \\
    \yThree^+ > \yOne^+ \land \yThree^+ > \yTwo^+ & \rightarrow \af^+ = \Amax~\land\\
    \omit{\rlap{$\xf + L \leq \xl \land \clock^+=t \land \clock \leq t \leq \clock+T \land \constctx$}}
    \end{align*}
    \vspace*{-1cm}
    \caption{Context assumptions for simplification}
    \label{fig:nnCtxt}
\end{wrapfigure}
Unfortunately, \Cref{eq:safe_nn:versaille_condition} cannot effectively be used for the NN verification directly as the NNs do not set the ego-cars acceleration ($\af^+$) but rely on surrounding software which computes $\af^+$ based of $\yOne,\yTwo,\yThree$ (and resets the clock variable $\clock$).
Moreover, $\af,\al,t$ and $\clock$ are no inputs to the NN and would thus need to be quantified over.
To make our verification condition practical, we derive a simplified version that we prove equivalent to \Cref{eq:safe_nn:versaille_condition}.
To this end, we begin by axiomatizing our assumptions on the NN's surroundings.
We assume the software correctly assigns $\af$ based on $\yOne,\yTwo,\yThree$, correctly manages clock variables and that we drive behind the other car (as mentioned above, we focus on this case).
We also set $\Amin=\Amax$ (as done in the official ABZ specificaton~\cite{ABZSpec}) and assume the known ranges of constants as formalized in \Cref{fig:nnCtxt}. %
Assuming $\modelplexCtxSym$, the system's invariant can then be simplified as follows:
\begin{align*}
\invariantCondSimpSym~\defeq&~0 \leq \vl \leq V \,\land\, 0 \leq \vf \leq V
\,\land\,%
                            \xf + L \leq \xl\,\land\,
                            \dist_\indEgo(\Bmin) + L < \dist_\indOther
\end{align*}
The simplified invariant makes sense intuitively as it matches the initial condition constraints in \Cref{eq:safety:back} on the variables in \nnInputVec.
Similarly, we simplify $\modelplexCondOrigSym$ by removing cases irrelevant to the ego-car driving behind, the management of clock variables and explicit mentions of $\af^+$.
This yields a simplified formula $\modelplexCondSimp$ (see \Cref{app:fig}).
For these simplifications, we prove equivalence to \Cref{eq:safe_nn:versaille_condition} in \KeYmaeraX{} under the assumption of \modelplexCtxSym:

\begin{lemma}[Simplified NN Verification] %
\label{lem:simp}
The following formula is valid:
\begin{align*}
&\modelplexCtx \rightarrow\\
&\Bigg(
\underbrace{
\left(
\begin{array}{c}
      \invariantCondSimpIn \rightarrow \\
      \modelplexCondSimp
\end{array}
\right)
}_{\text{simplified}}
\leftrightarrow
\underbrace{
\left(
\begin{array}{rl}
      \forall \af \forall \al&\Big(\left(\invariantCondOrigIn\right) \\
      &\rightarrow \modelplexCondOrigInOut\Big)
\end{array}
\right)
}_{\Cref{eq:safe_nn:versaille_condition}}
\Bigg)
\end{align*}
\end{lemma}
\looseness=-1
This serves as justification for verifying the simplified condition
$
\nnSpecSimp\enspace\defeq\enspace
\invariantCondSimpSym\left(\overline{\texttt{in}}\right) \rightarrow \modelplexCondSimpSym\left(\overline{\texttt{in}},\overline{\texttt{out}}\right)
$
on our NNs as we can assume $\modelplexCtxSym$. %
While $\nnSpecSimp$ is free of quantifiers, it still contains polynomial arithmetic (e.g.\ in $\dist_\indEgo(\Bmin)$).
In addition to the two cars modelled in \dL, the NN controller gets as input the states of up to three more cars (we will call these cars car 1 to car 5 with car 1 being the ego car).
For the single-lane case, the ego car's influence on crashes with cars 3-5 is very limited.
However, we know that car 2 can avoid a crash with car 3 if the velocity of car 3 is larger than the velocity of car 2 (e.g.\ by performing an emergency brake).
For now, we thus assume that for the extra cars $3 \leq i \leq 5$ it is guaranteed that car $i-1$ is slower than car $i$.
We thus encode these additional constraints on the state of cars 3-5 in a predicate $\nnSpecAdditional$ (see \Cref{app:fig}) and then verify the NN w.r.t.\ to the specification $\nnSpecAdditional \rightarrow \nnSpecSimp$.
$\nnSpecAdditional$ also contains constraints on the encoding of (non-)presence of cars and the NN's input space normalisation described in ABZ's specification~\cite{ABZSpec}.
In \Cref{sec:verification_results} we will see concrete examples for verifying NNs with respect to the full specification $\nnSpecAdditional \rightarrow \nnSpecSimp$, but we will first demonstrate that similar formulas can also be used for monitoring and shielding.

\paragraph{Justified Speculative Control and VeriPhy.}
\looseness-1
Assuming \modelplexCtxSym{} and \invariantCondSimpSym, it also holds that 
$
\modelplexCondSimp
\leftrightarrow
\modelplexCondOrigInOut
$.
Consequently, we can use the simplified monitoring condition not only for verification, but also for the construction of shields (JSC) and runtime monitors (VeriPhy).
JSC is meant to only check the runtime monitor when the observed behaviour matches the model.
To this end, JSC usually has a model monitor that checks whether a given state transition is explainable by the \dL environment model.
However, early experiments showed divergence in the simulation's environment and the \dL environment model which would effectively deactivate JSC in most of the state space (these observations will be discussed in more detail in the latter sections).
Consequently, we relaxed the model monitor to its ``most'' safety-critical parts and only check whether a given state is inside the invariant state space.

\looseness=-1
In this section, we have seen how ModelPlex conditions and invariants can
be applied even if they contain unobservable variables~\cite{DBLP:journals/corr/abs-1811-06502}.
This allows us to apply \dL-based monitoring, shielding and verification techniques independently of whether some variables (e.g.\ time or effective acceleration) are measurable or not.

\section{Verification Results} %
\label{sec:verification_results}
\begin{wraptable}[5]{r}{0.5\textwidth}
    \centering
    \vspace*{-1.2cm}
    \caption{Results of NN verification.}
    \begin{tabular}{c|l|l|l}
        \multirow{2}{*}{\textbf{NN}} & \multirow{2}{*}{\textbf{Time}}&\multicolumn{2}{c}{\# Crashes}\\\cline{3-4}
         & & \classicEnv{} & \frontBrake{} \\\hline\hline
        \Cref{subsec:model2sim:first} & 3.6 h & \hphantom{0,}538 & 3,593 \\\hline
        \Cref{subsec:model2sim:improved} & 1.9 h & 4,852 & 8,713\\
    \end{tabular}
    \label{tab:verification_table}
\end{wraptable}
\looseness=-1
A manual analysis of the agents in ABZ's case study~\cite{ABZSpec} uncovered that the agents' action space is different from its formal specification~\cite{ABZSpec}:
The \texttt{highway-env} simulator configuration admits different action spaces.
The provided agents used \texttt{DiscreteMetaAction} configuring the agent's action space as decreasing/increasing a \emph{reference velocity} $v_r\in\left\{0,5,\dots,35,40\right\}$ achieved via a low-level proportional controller.
This can lead to very different action outcomes compared to the specified action space.
For example, the \emph{brake} action as described in the formal specification \emph{always} leads to a deceleration (unless $\vf$ is zero already). %
In contrast, the ``brake'' action with the \texttt{DiscreteMetaAction} configuration can even lead to an \emph{acceleration} (e.g.\ if $\vf=10$ and $v_r=20$, ``braking'' sets $v_r$ to $15$ and the proportional controller accelerates so that $\vf=15$ is reached).
The safety guarantees and verification conditions derived in \Cref{sec:model,sec:safe_nn} thus only apply to the written specification, but not to the simulator's default configuration violating its own description.
We trained a new NN using \texttt{highway-env}'s \texttt{DiscreteAction} configuration option\footnotemark{} (otherwise using the default configuration) 
that can brake, idle or accelerate \emph{directly} ($\af\in\left\{\Bmax,0,\Amax\right\}$).
\footnotetext{Python's weak type system makes the configuration especially error-prone: 
The \texttt{acceleration\_range} is configured via a \emph{2-tuple} $(\texttt{min},\texttt{max})$.
Accidentally providing a \emph{list} of actions interpolates discrete accelerations between the list's first two elements.}
We discuss verification of this NN (\Cref{subsec:model2sim:first}) and an improved version (\Cref{subsec:model2sim:improved}).
Results are reproducible via our artifact~\cite{zenodoArtifact}.

\subsection{A First Attempt at Verification}
\label{subsec:model2sim:first}
\begin{wrapfigure}[8]{r}{0.5\linewidth}
    \centering
    \vspace*{-0.85cm}
    \includegraphics[width=\linewidth]{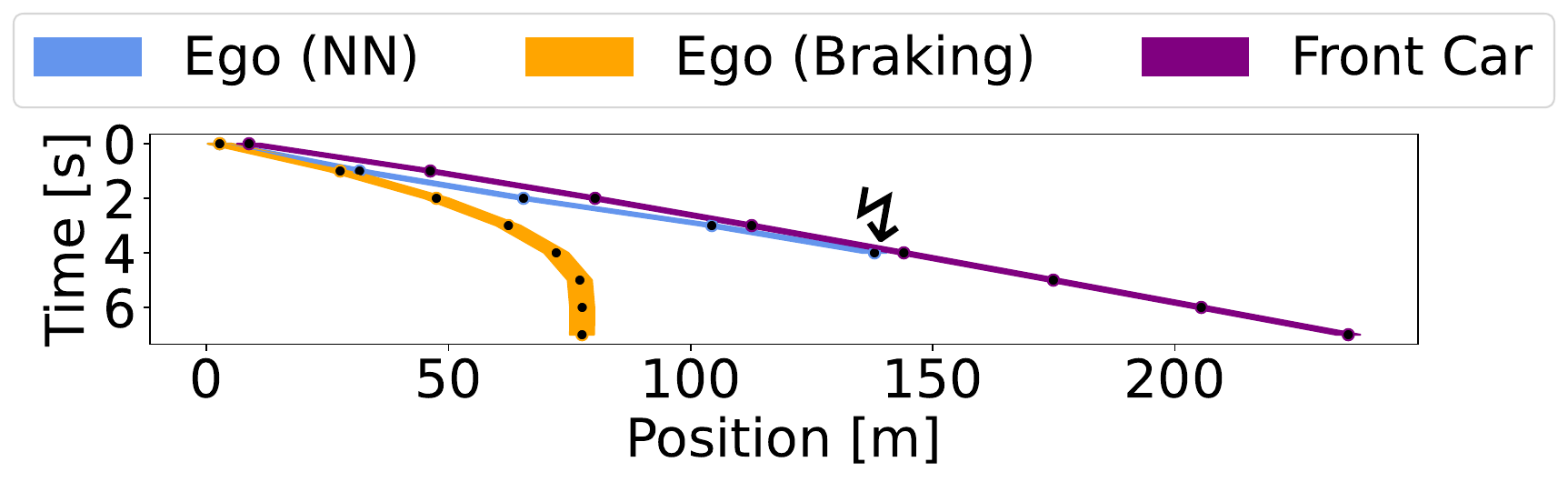}
    \vspace*{-1cm}
    \caption{One of 538 examples of unsafe NN behaviour in \classicEnv{} environment (x-axis shows position, y-axis shows time). Braking could have avoided a crash ($\lightning$).}
    \label{fig:crash-1}  
\end{wrapfigure}
\looseness=-1
As a first step we attempted to verify the NN for two cars w.r.t.\ the specification derived in \Cref{sec:safe_nn} using 
NCubeV~\cite{TeuberVerSAILLE2024}
which supports polynomial arithmetic specifications.
In case a specification cannot be proven, NCubeV is also capable of enumerating \emph{all} counterexample regions (represented as polytopes) to a given specification.
Notably, successful verification would, by construction, guarantee that the two cars on the highway will never crash -- independently of trip time.
The trained NN instead turned out to be unsafe:
NCubeV returned \textbf{14,917 counterexample regions}.
Computing these counterexamples took 3.6 hours (see \Cref{tab:verification_table}).
However, verifying safety is often quicker than enumerating all counterexamples for NCubeV~\cite{TeuberVerSAILLE2024}.
Each counterexample region has a representative input violating the specification.
These inputs can be used to sample trajectories from the simulator to find concrete crashes.
\Cref{fig:crash-1} shows one of 538 concrete crashes we observed when the front car is controlled by the \emph{Intelligent Driver Model}~\cite{treiber2000congested} (IDM).
Importantly, these crashes could have all been avoided by braking.
When the front car is configured to perform an emergency brake, our sampling strategy yielded 3,593 crash trajectories.

\looseness=-1
These observations raise two questions:
\begin{inparaenum}
    \item Why did the NN not learn to brake in time?
    \item Is there nonetheless a way to safely deploy the NN at hand?
\end{inparaenum}
One answer to the former question can be found in the IDM.
While originally derived as a means to understand traffic congestion, \texttt{highway-env} uses the model to control the environment's cars.
Due to the way IDM is set up, the ego-car rarely experiences emergency brakes of front cars and thus does not learn to account for them (as indicated by over 3,000 crash trajectories for emergency braking front cars).
The \texttt{highway-env} simulation environment is thus another example of a previously observed phenomenon that worst-case scenarios which occur with low probabilities during training are typically not learned by reinforcement learning agents and that these errors can be uncovered by formal verification~\cite{TeuberVerSAILLE2024}.
In the present environment, this issue is exacerbated by the fact that the agent learns that it can brake with acceleration $\af=\Bmax$ although (according to the specification) the acceleration can be as little as $\Bmin$.

\looseness=-1
We now turn to the question of how the NN can be safeguarded under the given conditions.
To this end, we evaluated the NN's empirical performance (reward) and crash behaviour w.r.t.\ the IDM front-car (\classicEnv{}) as well as w.r.t.\ an environment where the front car performs emergency brakes and the ego car can only decelerate with $\Bmin$ (\frontBrake{}).
We evaluate the stand-alone NN, a monitored version using a Python implementation of the VeriPhy approach (this implementation comes without the rigorous compilation guarantees of VeriPhy~\cite{Bohrer2018}) and a shield for the NN using JSC~\cite{Fulton2018}.
We evaluate w.r.t.\ initial conditions satisfying the invariant over 1000 sampled trajectories\footnote{
Initial conditions are generated via rejection sampling.
For a sufficiently high success rate, we had to reduce the simulator's car density parameter.}.
The results are in \Cref{tab:comparison_veriphy_jsc_inside_inv} and indicate relatively consistent behaviour w.r.t.\ to the reward standard deviation.
Empirically, we observe that the agent trained w.r.t.\ to the IDM model (\classicEnv{} environment) crashes in 996 out of 1000 cases when evaluated w.r.t.\ a braking front car (\frontBrake{} environment).
Our investigation indicates that the dynamics in \classicEnv{} lack diversity in at least three dimensions:
First, the environment assumes maximal braking power
(contradicting the formal specification~\cite{ABZSpec});
Secondly, the environment very rarely simulates braking front cars.
Finally, we posit that \classicEnv{} only samples initial conditions from a small subset of admissible initial conditions as our verifier found many concrete initial conditions that lead to crashes in \classicEnv{}.
Importantly, VeriPhy and JSC allow us to (provably!) avoid these crashes by intervening when the model chooses unsafe actions.
We observe that, based on the reward function, JSC matches the best results across both environments while leading to 0 crashes.
VeriPhy's and JSC's behaviour differs in their statistics on taken actions:
For example, JSC chooses the idle action in 6.3\% of time steps while VeriPhy never chooses this action.

\begin{table}[t]
    \centering
    \caption{\textbf{Empirical results} for the original, monitored (VeriPhy) and shielded (JSC) controller given initial conditions \emph{inside} the safely controllable (i.e.\ invariant) state space. The velocity bounds of JSC's invariant check had to be modified as the simulator occasionally produces velocities outside $[0,V]$ which would otherwise deactivate JSC.}
    \begin{tabularx}{\textwidth}{X || r | r | r | r | r | r}
        \multirow{2}{*}{Env} &
        \multicolumn{2}{c|}{Original NN} &
        \multicolumn{2}{c|}{VeriPhy} &
        \multicolumn{2}{c}{JSC*}\\\cline{2-7}
        & \multicolumn{1}{c|}{Reward} & \multicolumn{1}{c|}{Crash}
        & \multicolumn{1}{c|}{Reward} & \multicolumn{1}{c|}{Crash}
        & \multicolumn{1}{c|}{Reward} & \multicolumn{1}{c}{Crash}\\\hline\hline
        {\classicEnv{} (IDM)}
        & $\mathbf{17.63} \pm 0.21$ & \textbf{0}\hphantom{.0}\%
        & $16.72 \pm 0.32$ & \textbf{0}\hphantom{.0}\%
        & $\mathbf{17.63} \pm 0.21$ & \textbf{0}\hphantom{.0}\%\\
        \frontBrake{}
        & ${5.44} \pm 1.27$ & 99.6\%
        & $\mathbf{16.47} \pm 0.05$ & \textbf{0}\hphantom{.0}\%
        & $\mathbf{16.47} \pm 0.05$ & \textbf{0}\hphantom{.0}\%
    \end{tabularx}
    \label{tab:comparison_veriphy_jsc_inside_inv}
\end{table}

\subsection{An Improved NN Controller}
Based on the results from \Cref{subsec:model2sim:first} we attempted to train a second agent.
To this end, we also modified the training.
First, we enforce that 80\% of initial states satisfy the invariant
(like above, we achieve this by sampling with reduced car density).
Second, we modified the behaviour of environment variables:
In each control round a car initiates (and then continues in subsequent steps) an emergency brake with 15\% probability.
Our objective is to increase the likelihood of the agent experiencing worst-case behaviour of the environment as a learning opportunity -- especially in situations where crashes can be avoided.
Finally, as NN verification and counterexample region enumeration scales exponentially with the NN's size, we reduce the NN to two layers with 16 neurons each.
Unlike the provided environment (20k steps) we train for up to 40k steps and choose the best-performing model (achieved after 22k steps).
To simplify the task, we furthermore assume $\Bmin=\Bmax=-5.0$.
An evaluation across 1,000 initial conditions for \frontBrake{} (with $\Bmin=-5$) yielded a reward of $16.08 \pm 0.07$ with $0$ crashes.
Compared to the first NN's performance for the \frontBrake{} environment in \Cref{tab:comparison_veriphy_jsc_inside_inv} this is a notable performance improvement.
Given these promising results, we attempted verification w.r.t.\ the full NN specification (2 to 5 cars).

\label{subsec:model2sim:improved}
\begin{wrapfigure}[13]{r}{0.275\textwidth}
    \centering
    \vspace*{-2em}
    \includegraphics[width=0.92\linewidth]{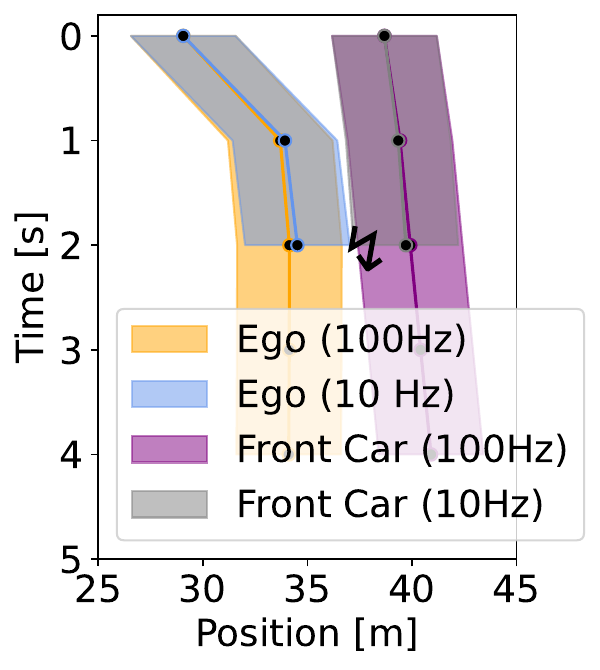}
    \vspace*{-0.5em}
    \caption{An \emph{Euler Crash} ($\lightning$): %
    Occurrence depends on Euler approximation resolution.}
    \label{fig:euler-crash}
\end{wrapfigure}
\looseness=-1
Verification took 1.9 hours and still returned \textbf{11,059 counterexample regions}.
Simulations with the representative inputs for the returned regions uncovered
4852 crashes in the \classicEnv{} simulation (using IDM) and 8713 crashes in the \frontBrake{} simulation (with $\Bmax=\Bmin=-5$; see \Cref{tab:verification_table}).
Surprisingly, for the two simulations we resp. found 181 and 40 cases that even produced a crash when the ego-car performed an emergency brake!
This is surprising as our \dL proof states that braking should keep our system safe.
A closer examination uncovered that these are \emph{Euler Crashes}, i.e.\ the occurrence of a crash depends on the resolution of the Euler approximation.
For a finer step size of the Euler approximation, the spurious crash disappears.
Importantly, in almost all cases the crash produced by the NN remained. 
An example for an Euler Crash
(evaluated with 10 and 100 Euler steps per second of evolution) can be found in \Cref{fig:euler-crash}.

\section{The Model2Simulation Gap}
\looseness-1
Overall, this work has not only derived an abstract \dL model, but also demonstrated in practice that verification can serve as a powerful tool to detect flaws in reinforcement learning systems.
Across two NNs our analysis uncovered numerous concrete counterexamples for NNs even though they performed \emph{flawlessly} in their respective simulations.
Throughout, we attempted to trace these faults to design choices in the simulator such as the intelligent driver model or the sampling method for choosing initial conditions.
Overall, our results provide strong evidence that \emph{as is} the \texttt{highway-env} simulator provides no reliable basis for the training of safe car control NNs.
However, we believe the detected issues point to a larger issue concerning inconsistencies between models and simulators in general.
While we consistently took the stance that our model is correct and the simulation is to blame, in reality, this is not always the case:
For example, was it justified that we changed the NN's action space or should we have built an entirely different \KeYmaeraX model?
Here, we believe our choice was justified by ABZ's specification document~\cite{ABZSpec}, but such documentation may not always be available.
While this work demonstrates how far \dL-based safety certification for NN Control Systems has come, it also underscores the intricate issues of interlinking simulation-based evaluation with a symbolic, \dL-based analysis.

\subsubsection*{Acknowledgements.}
This work was supported by funding from the pilot program Core-Informatics of the Helmholtz Association (HGF) and by an Alexander von Humboldt Professorship.

\bibliographystyle{splncs04}
\bibliography{main, platzer}

\clearpage
\appendix
\section{Additional Figures}
\label{app:fig}

\begin{figure}
    \centering
\begin{alignat}{2}\label{eq:safefront}
    \xl &+ L \leq \xf \land \big(\Amin \leq \af \land\fram{\dist}_\indOther + L < \fram{\dist}_\indEgo(\Amin) \nonumber\\
& \lor \af\leq \Amin \land \fram{v}_{\indEgo} + \af T > 0 \land \fram{\dist}_\indOther + L < \fram{\dist}_\indEgo(\af)
\nonumber\\
    & \lor \af \leq \Amin \land \fram{v}_{\indEgo} + \af T \leq 0 \land \fram{\dist}_\indOther + L < \fram{\dist}_\indEgo(\Amin) + (\frac{-\af}{\Amin} + 1)(\frac{\af}{2}T^2 + \fram{v}_{\indEgo}T)
    \big) \nonumber
\end{alignat}
    \caption{$\safeFrontT$}
\end{figure}
\begin{table}
    \centering
    \caption{Verification using \KeYmaeraX: the number of steps is the size of the implicit proof tree from the kernel including lemmas. QE time is included in the total duration.}
    \begin{tabular}{l||l|l|l|l|l}
        Proof & Status & Tactic Size & Duration [ms] & QE [ms] & Steps\\
        \hline\hline
        \safeBackT{} invariant of \dyn{} & proved & 48 & 21,362 & 10,648 & 15,378\\
        \safeFrontT{} invariant of \dyn{} & proved & 97 & 14,471 & 9,858 & 17,456\\
        $\model(\ctrlT)$ - Back~(\ref{eq:safety:back}) & proved & 84 & 14,701 & 8,241 & 21,068\\
        $\model(\ctrlT)$ - Front~(\ref{eq:safety:front}) & proved & 110 & 16,407 & 6,865 & 25,689\\
        \hline
        Controllers Refinement & proved & 143 & 4,093 & 1,875 & 2,446\\
        Models Refinement~(\ref{eq:refinement}) & proved & 88 & 2,878 & 0 & 3,716\\
        $\modelNN$ - Back & proved & 8 & 1,249 & 0 & 24942\\
        \hline
        ModexPlex simp (\Cref{lem:simp}) & proved & 326 & 104,634 & 70,533 & 11902\\
    \end{tabular}
    \label{tab:kyx}
\end{table}

\begin{figure}
    \centering
    \begin{alignat*}{2}
    y_1^+ &\geq y_2^+ \land&& y_1^+ \geq y_3^+\\
    &\lor y_2^+ >&&\, y_1^+ \land y_2^+ \geq y_3^+ \land \\
    & &&\big(
    \Bmin \leq 0 \leq \Amax \land \vf \geq 0 \land {\dist}_{\indEgo}(\Bmin) + (\frac{0}{\Bmin} + 1)T\vf + L < \dist_{\indOther}
    \big)\\
    & \lor
    y_3^+ >&&\,y_1^+ \land y_3^+ > y_2^+ \land \big(
    \Bmax \leq \Amax \leq \Bmin \land \dist_{\indEgo}(\Bmin) + L < \dist_{\indOther}\\
    & &&\lor
    \Bmin \leq \Amax \land \vf + \Amax T < 0 \land \dist_{\indEgo}(\Amax) + L < \dist_{\indOther}\\
    & &&\lor
    \Bmin \leq \Amax \land \vf + \Amax T \geq 0 \land \dist_{\indEgo}(\Bmin) + (\frac{-\Amax}{\Bmin} + 1)(\frac{\Amax}{2}T^2 + T\vf) + L < \dist_\indOther
    \big)
\end{alignat*}

    \caption{$\modelplexCondSimp$}
\end{figure}

\begin{figure}
    \centering
\begin{alignat*}{5}
&p_{\indEgo} = 1 &\,\land\, 0 \leq \frac{\xf}{5*V} \leq 1 &\,\land\,0 \leq \frac{\vf}{2*V} \leq 1 &\,\land\,y_{\indEgo}=0 &\,\land\,w_{\indEgo}=0  \,\land\\
&p_{\indOther} = 1 &\,\land\,  -1 \leq \frac{\xl-\xf}{5*V} \leq 1 &\,\land\, 0 \leq \frac{\vl-\vf}{2*V} \leq 1  &\,\land\, y_{\indOther} = 0  &\,\land\,  w_{\indOther} = 0 \,\land\\
&0 \leq p_3 \leq 1 &\,\land\, -1 \leq \frac{x_3-\xf}{5*V} \leq 1 & \,\land\, 
0 \leq \frac{v_3-\vf}{2*V} \leq 1 &\,\land\, 
y_3 = 0 & \,\land\,  w_3 = 0  \,\land\\
&0 \leq p_4 \leq 1 &\,\land\, -1 \leq \frac{x_4-\xf}{5*V} \leq 1 & \,\land\, 
0 \leq \frac{v_4-\vf}{2*V} \leq 1 &\,\land\, 
y_4 = 0 & \,\land\,  w_4 = 0  \,\land\\
&0 \leq p_5 \leq 1 &\,\land\, -1 \leq \frac{x_5-\xf}{5*V} \leq 1 & \,\land\, 
0 \leq \frac{v_5-\vf}{2*V} \leq 1 &\,\land\, 
y_5 = 0 & \,\land\,  w_5 = 0  \,\land\\
\omit{\rlap{$-1000 \leq \yOne \leq 1000\,\land\,-1000 \leq \yTwo \leq 1000\,\land\,-1000 \leq \yThree \leq 1000\,\land$}}\\
\omit{\rlap{$ \left( p_3 = 0 \lor p_3 = 1\right) \,\land\, \left( p_4 = 0 \lor p_4 = 1\right) \,\land\, \left( p_5 = 0 \lor p_5 = 1\right) \,\land\,$}}\\
\omit{\rlap{$ \left(p_3 = 0 \rightarrow \left( x_3 = 0 \land v_3 = 0 \right) \right)\,\land\,$}}\\
\omit{\rlap{$
\left(p_4 = 0 \rightarrow \left( x_4 = 0 \land v_4 = 0 \right) \right) \,\land\,$}}\\
\omit{\rlap{$
\left(p_5 = 0 \rightarrow \left( x_5 = 0 \land v_5 = 0 \right) \right) \,\land\,
$}}\\
\omit{\rlap{$ \left(p_3 = 1 \rightarrow \left( \xl + L \leq x_3 \land \vl \leq v_3 \right) \right)\,\land$}}\\
\omit{\rlap{$ \left(p_4 = 1 \rightarrow \left( x_3 + L \leq x_4 \land v_3 \leq v_4 \,\land\, p_3 = 1 \right) \right)\,\land$}}\\
\omit{\rlap{$ \left(p_5 = 1 \rightarrow \left( x_4 + L \leq x_5 \land v_4 \leq v_5 \,\land\, p_4 = 1 \right) \right)$}}
\end{alignat*}
    \caption{$\nnSpecAdditional$: $p$ is the presence indicator, $y$ the latitudinal position and $w$ the latitudinal velocity. Additionally, we must configure the verifier so that the inputs are considered w.r.t.\ their normalized value, e.g.\ the ego car position input has to be $\frac{\xf}{5V}$.
    Moreover, the other car inputs are relative to the front car, e.g.\ the front car position is $\frac{\xl-\xf}{5 V}$.
    Finally, we instantiate the constants ($V=40,T=1,L=5,\Bmax=-5,\Amax=5,\Amin=5$ and depending on query $\Bmin\in\left\{-3,-5\right\}$).}
\end{figure}
\end{document}